\documentclass[a4paper]{aa}
\usepackage[utf8]{inputenc}
\usepackage{graphics}
\usepackage{graphicx}
\usepackage{amsmath}
\usepackage{amssymb}
\usepackage{color}
\usepackage{xcolor}
\usepackage{threeparttable,booktabs}
\usepackage{natbib,twoopt}
\usepackage{lipsum}
\definecolor{CornflowerBlue}{rgb}{0.39,0.58,0.93}
\usepackage[breaklinks=true,colorlinks=true,linkcolor=blue,citecolor=CornflowerBlue,linkbordercolor=blue,citebordercolor=magenta,urlbordercolor=green]{hyperref}

\title{Projection effects in galaxy cluster samples: \\ insights from X-ray redshifts}

\author{M. E. Ramos-Ceja\inst{1} \and F. Pacaud\inst{1} \and T. H. Reiprich\inst{1} \and K. Migkas\inst{1} \and L. Lovisari\inst{2} \and G. Schellenberger\inst{2}}

\institute{
 Argelander-Institut f{\"u}r Astronomie, Universit{\"a}t Bonn, Auf dem H{\"u}gel 71, 53121 Bonn, Germany \\
 \email{miriam@astro.uni-bonn.de}
 \and 
 Center for Astrophysics | Harvard \& Smithsonian, 60 Garden Street, Cambridge, MA 02138, USA}

\date{Accepted 2019}

\abstract
{
Up to now, the largest sample of galaxy clusters selected in X-rays comes from the ROSAT All-Sky Survey (RASS). Although there have been many interesting clusters discovered with the RASS data, the broad point spread function (PSF) of the ROSAT satellite limits the amount of spatial information of the detected objects. This leads to the discovery of new cluster features when a re-observation is performed with higher resolution X-ray satellites. Here we present the results from {\it XMM-Newton} observations of three clusters: RXCJ2306.6-1319, ZwCl1665 and RXCJ0034.6-0208, for which the observations reveal a double or triple system of extended components. These clusters belong to the extremely expanded HIghest X-ray FLUx Galaxy Cluster Sample (eeHIFLUGCS), which is a flux-limited cluster sample ($f_\textrm{X,500}\geq 5\times10^{-12}$~erg~s$^{-1}$~cm$^{-2}$ in the $0.1-2.4$~keV energy band). For each structure in each cluster, we determine the redshift with the X-ray spectrum and find that the components are not part of the same cluster. This is confirmed by an optical spectroscopic analysis of the galaxy members. Therefore, the total number of clusters is actually 7 and not 3. We derive global cluster properties of each extended component. We compare the measured properties to lower-redshift group samples, and find a good agreement. Our flux measurements reveal that only one component of the ZwCl1665 cluster has a flux above the eeHIFLUGCS limit, while the other clusters will no longer be part of the sample. These examples demonstrate that cluster-cluster projections can bias X-ray cluster catalogues and that with high-resolution X-ray follow-up this bias can be corrected.
}


\makeatletter
\renewcommand*\aa@pageof{, page \thepage{} of \pageref*{LastPage}}
\makeatother

\begin{document}

\maketitle


\section{Introduction}

Systematic searches for galaxy clusters have traditionally been conducted at optical wavelengths \citep[e.g.,][]{Abell1958}, the main selection criterion being the statistical excess of cluster galaxies with respect to the {\it background} along the line of sight. While some optical catalogues still are the largest compilations of galaxy clusters \citep[e.g.,][]{Wen2012}, the inherent biases to the optical selection process, most important projection effects, lead to false detections that are difficult to correct for in statistical studies of cluster properties \citep[e.g.,][]{Costanzi2019}.

In the last two decades X-ray and sub-millimetre observations have provided some of the purest galaxy cluster catalogues \citep[e.g.,][]{Boehringer2000,Boehringer2001,Planck2016}. Although these samples tend to detect a particular type of clusters \citep[see][for a description of several biases]{Giodini2013}, they overcome most of the optical biases. Especially, X-rays provide powerful means of selecting galaxy clusters and characterizing their properties. This is possible given that the X-ray flux is proportional to the square of the electron density of the intra-cluster medium (ICM), and therefore it is less prone to line of sight structure projection. Furthermore, the gas only gets hot enough to emit X-rays if it sits in a deep gravitational potential. Last but not least, in X-rays, clusters are identified as single objects, basically a huge hot plasma cloud, and not just as the sum of individual galaxies. X-ray observations of galaxy clusters thus allow the compilation of cluster samples that are almost unaffected by projection effects and have very few false detections. Finally, the analysis of the X-ray emission spectrum of the ICM also allows the determination of various cluster properties such as e.g. the gas mass, temperature, metal abundance, and the redshift of the cluster. The ICM is a plasma that can generally assumed to be optically thin in collisional equilibrium, whose spectrum is described by a bremsstrahlung spectrum together with emission lines of highly ionized metals \citep[e.g.,][]{Sarazin1986,Peterson2006,Boehringer2010}.

Cluster redshifts and total masses are crucial measurements for cosmological studies based on galaxy clusters. The cluster redshift information is usually obtained from optical spectroscopic follow-up observations of the cluster member galaxies, however, this process is limited to one or few cluster member galaxies and it can require long observational campaigns for large cluster samples. While these spectroscopic measurements give very precise cluster redshifts ($\Delta z/(1+z)<0.01$), the direct use of X-ray data also yield complementary redshift estimations \citep[e.g.,][]{Yu2011,Lloyd-Davies2011,Liu2018}.

\begin{table*}[ht]
\caption{Cluster properties extracted from the MCXC catalogue. The luminosity is given in the $0.1-2.4$~keV energy band.}
\centering
\begin{threeparttable}
\renewcommand{\arraystretch}{1.1} 
  \begin{tabular}{l c c c c c c}
   \hline
   \hline
    Cluster & R.A. & Dec. & $z$ & $r_{500}$ & $L_{500}$ & Parent\\
     & (J2000) & (J2000) & & [kpc] & [$10^{43}$~erg~s$^{-1}$] & catalogue\\
   \hline
     RXCJ2306.6-1319 & $346.650$ & $-13.320$ & $0.066$ & $772$ & $5.7$ & REFLEX\\
     ZwCl1665 & $125.798$ & $+4.356$ & $0.029$ & $630$ & $1.9$ & BCS\\
     RXCJ0034.6-0208 & $8.650$ & $-2.140$ & $0.081$ & $897$ & $12.5$ & REFLEX\\
   \hline
   \hline
  \end{tabular}
\label{tab:knownproperties}
\end{threeparttable}
\end{table*}

X-ray cluster catalogues based on the ROSAT All-Sky Survey \citep[RASS,][]{Truemper1992,Truemper1993}, such as the Northern ROSAT All-Sky Survey \citep[NORAS,][]{Boehringer2000}, the Bright Cluster Sample \citep[BCS,][]{Ebeling1998}, or the ROSAT-ESO Flux Limited X-ray Galaxy Cluster Survey \citep[REFLEX,][]{Boehringer2001} typically consist of a few hundred clusters. However, due to the limited RASS spatial resolution (of order $1$~arcmin) and the limited galaxy redshift follow-up, a fraction of the systems identified as single clusters may indeed be double clusters or even two unrelated clusters with a small projected separation. The {\it XMM-Newton} and {\it Chandra} observatories have larger effective area and higher angular resolution than ROSAT, which have made possible a deeper analysis of the ICM and help us to identify such double cluster systems. For example, upon close inspection with {\it XMM-Newton}, the cluster Abell~1644 turned out to be a merging double cluster \citep[][]{Reiprich2004}. Moreover, the substructure selection has been an important issue when using complete galaxy cluster samples for cosmological analysis; e.g., \citet[][]{Schellenberger2017} developed and employed an objective, automated substructure removal procedure, which works well with high-quality data.

In the near future, the extended ROentgen Survey with an Imaging Telescope Array \citep[eROSITA,][]{Predehl2010,Merloni2012} is expected to detect $\sim10^5$ clusters \citep[][]{Pillepich2012,Pillepich2018,Clerc2018}. The cosmological studies that will be carried out with these detected clusters will require the knowledge of the cluster redshift. For eROSITA, this will be approached in different ways: lower-precision photometric redshifts are readily available through large area multicolor surveys like DES \citep[][]{DES2016}, ATLAS \citep[][]{Shanks2015}, PanSTARRS \citep{Panstarrs2016}. Special projects to acquire higher-precision spectroscopic redshifts are underway \citep[e.g., 4MOST and SPIDERS,][]{4most2014,Clerc2016,Zhang2017}. Additionally, low-precision ($\Delta z / (1+z)<0.1$) X-ray redshifts will be directly available from the eROSITA survey data for a subsample of clusters \citep{Borm2014}. It is also expected that a subsample, of order $\sim10^3$ clusters, will have high-resolution data from {\it XMM-Newton}, {\it Chandra} and XRISM \citep{Tashiro2018}.

In this paper, we report on the {\it XMM-Newton} observations of three clusters detected in the RASS, for which the {\it XMM-Newton} data reveals a double or even triple X-ray morphology. We use X-ray and optical spectroscopy information and find that the clusters detected in the RASS data are not only not single clusters but even just projection effects and the double X-ray morphologies are indeed separated clusters at different redshifts. Throughout this paper, we assume a flat $\Lambda$CDM cosmology with $\Omega_\textrm{m}=0.3$ and $H_0=70~\rm km~s^{-1}~Mpc^{-1}$.


\section{Selected clusters}
\label{sect:knowninfo}

The extremely expanded HIghest X-ray FLUx Galaxy Cluster Sample \citep[eeHIFLUGCS,][]{Reiprich2017} comprises all clusters with $f_{\textrm{X,500}}\geq5\times10^{-12}$~erg~s$^{-1}$~cm$^{-2}$ in the $0.1-2.4$~keV energy band (almost $400$ clusters). This selection takes into account only RASS-based catalogues, whose homogenized fluxes are calculated from the luminosities determined in the Meta-Catalog of X-Ray Detected Clusters of Galaxies \citep[MCXC,][]{Piffaretti2011}. As of October 2018, $\sim87\%$ of the sample ($\sim330$ clusters) has good quality data in either {\it XMM-Newton} and/or {\it Chandra}. Using images in the $0.5-2$~keV energy band we perform a visual inspection of the good {\it XMM-Newton} observations ($\sim 240$ clusters) and identify that approximately $15$ clusters show two or more clear extended structures, i.e. objects with diameters larger than $2$~arcmin. These objects are either merging systems or uncorrelated objects along the line of sight. In this work, we present $3$ clusters that fall into the latter category. In this work we present the X-ray spectral analysis that allows us to determine that the extended objects present in these observations have a different redshift from the one reported in MCXC and distinct among themselves.

In the following, we present the previously known X-ray information of the three clusters. Table~\ref{tab:knownproperties} shows a summary of the most important known features \citep[from][]{Piffaretti2011}.

\begin{table*}[ht]
\caption{Observation ID, unfiltered time of pn, clean exposure time of each EPIC cameras, IN/OUT ratio, and hydrogen column density value for the three observations used in this paper.}
 \centering
\begin{threeparttable}
\renewcommand{\arraystretch}{1.1} 
  \begin{tabular}{l c c c c c c c}
   \hline
   \hline
    Cluster & Obs. ID & Total & MOS1 & MOS2 & pn & IN/OUT ratio & $n_\textrm{H}$ \\
     &  & [ks] & [ks] & [ks] & [ks] & MOS1/MOS2/pn & [$10^{20}$~cm$^{-2}$]\\
   \hline
     RXCJ2306.6-1319 & $0765030201$ & $11.4$ & $8.1$ & $8.92$ & $5.8$ & $1.061/1.118/1.010$ & $3.0$ \\
     ZwCl1665 & $0741580501$ & $12.6$ & $10.5$ & $15.3$ & $9.9$ & $1.077/1.006/1.098$ & $2.5$ \\
     RXCJ0034.6-0208 & $0720250401$ & $11.2$ & -- & $7.3$ & $11.0$ & --$/1.001/0.961$ & $3.1$ \\
   \hline
   \hline
  \end{tabular}
\label{tab:XMMobs}
\end{threeparttable}
\end{table*}

\subsection{RXCJ2306.6-1319}

According to the MCXC, the cluster RXCJ2306.6-1319 (MCXC~J2306.5-1319) has been catalogued in REFLEX at $z=0.066$, with $r_{500}=0.772$~Mpc\footnote{$r_{500}$ is the radius within which the mean over-density of the galaxy cluster is $500$ times the critical density at the cluster redshift.} ($\sim10.17$~arcmin) and a luminosity $L_{500}=5.74\times10^{43}$~erg~s$^{-1}$ in the $0.1-2.4$~keV rest-frame energy band.

RXCJ2306.6-1319 is located $\sim9.6$~arcmin from another MCXC cluster, RXCJ2306.8-1324 (MCXC~J2306.8-1324). This cluster has been identified in the SGP \citep[A Catalog of Clusters of Galaxies in a Region of 1 Steradian around the South Galactic Pole,][]{Cruddace2002} at $z=0.066$, with $r_{500}=0.710$~Mpc ($\sim9.36$~arcmin) and $L_{500}=3.81\times10^{43}$~erg~s$^{-1}$ in the $0.1-2.4$~keV energy band. \cite{Boehringer2004} have a note on these clusters: RXCJ2306.6-1319 is one out of nine REFLEX clusters not listed in SGP, while RXCJ2306.8-1324 is one out of six clusters listed in the SGP with a flux above the REFLEX flux limit, but in REFLEX it has a flux lower than the REFLEX flux limit.

\subsection{ZwCl1665}

The MCXC cluster ZwCl1665 (MCXC~J0823.1+0421) has been catalogued in the BCS at $z=0.029$, with $r_{500}=0.630$~Mpc ($\sim17.88$~arcmin) and $L_{500}=1.92\times10^{43}$~erg~s$^{-1}$ in the $0.1-2.4$~keV energy band.

\subsection{RXCJ0034.6-0208}
\label{subsec:rxcj0034.6}

RXCJ0034.6-0208 (MCXC~J0034.6-0208) has been identified in the REFLEX catalogue at $z=0.081$, with $r_{500}=0.897$~Mpc ($\sim9.77$~arcmin) and $L_{500}=1.25\times10^{44}$~erg~s$^{-1}$ in the $0.1-2.4$~keV energy band. In the REFLEX catalogue there is an additional note about this cluster: ``DL, two maxima/E-W'', which we interpret as a double cluster displaying two X-ray maxima in the East-West direction. Moreover, RXCJ0034.6-0208 has been associated with one {\it Planck} cluster: PSZ2~G113.02-64.68 \citep{Planck2016}.

At a distance of $\sim9.6$~arcmin from RXCJ0034.6-0208, the MCXC cluster RXCJ0034.2-0204 (MCXC~J0034.2-0204) is located. This cluster has also been identified in the SGP catalogue at $z=0.082$, with $r_{500}=0.893$~Mpc ($\sim9.62$~arcmin) and $L_{500}=1.23\times10^{44}$~erg~s$^{-1}$ in the $0.1-2.4$~keV energy band. This cluster is also known as SH518 \citep{Shectman1985}.

The case of RXCJ0034.6-0208 is very similar to RXCJ2306.6-1319. According to \cite{Boehringer2004}, RXCJ0034.6-0208 is one out of nine REFLEX clusters not listed in SGP.


\section{{\bf \emph{XMM-Newton}} observations and data analysis}
\label{sect:obsdataan}

The three galaxy clusters described in Section~\ref{sect:knowninfo} have been observed by {\it XMM-Newton}. RXCJ0034.6-0208 has been observed four times with {\it XMM-Newton} (Obs. ID 0675470901, 0675472601, 0720250401, 0720252901), but three of them are entirely flared and cannot be used for a spectral analysis. Only Obs. ID 0720250401 (taken with the thin filter) has usable data, but only for the MOS2 and pn detectors. MOS1 presented some anomalies and it did not work in the necessary conditions to collect data. The observations for RXCJ2306.6-1319 and ZwCl1665 were taken with the medium filter (see Table~\ref{tab:XMMobs} for the Obs. ID), and all three observations were set to Full Frame mode. ZwCl1665 has also been observed with {\it Chandra} ($\sim10$~ks). We briefly discuss some results in Section~\ref{subsect:epzTZ}.

\subsection{Data reduction}

We retrieved the observation data files (ODFs) from the {\it XMM-Newton} archive and reprocessed them with the X{\small MMSAS} v16.0.0 software. We used the tasks {\tt emchain} and {\tt epchain} to generate calibrated event lists from the ODFs. 

We performed a filtering process to clean the data from periods of soft-proton flares. For this we created a light curve in the $0.3-10$~keV energy band with bins of $52$~s for MOS and $26$~s for pn, and selected only good events (by applying the expressions {\tt \#XMMEA\_EM} for MOS and {\tt (FLAG \& 0x766b0000)==0} for pn). Following \citet{Pratt2002}, we fitted a Poisson distribution to the light curve, and we applied a $\pm3\sigma$ thresholding. We rejected all time intervals exceeding this threshold and we obtained the good time intervals (GTIs). Table~\ref{tab:XMMobs} shows the clean exposure time of the observations.

Since a residual contribution from soft protons may still exist after the initial data filtering, we quantified this source of contamination for our data. For this, we determined the so-called IN/OUT ratio \citep{DeLuca2004,Leccardi2008}, which is the ratio of the surface brightness inside the field-of-view (FoV) to the one measured in the unexposed corners. The signal in the corners provide an estimate of the cosmic-ray induced instrumental background as focused X-ray photons or soft protons cannot pass through the filter support structure. On the other hand, performing the measurement at high energies ensures that the signal inside the FoV is also free from genuine X-ray photons due to the low telescope effective area. Therefore an excess of the IN/OUT ratio above unity indicates soft-proton contamination. \citet{DeLuca2004} have shown that IN/OUT ratios~$<1.15$ indicate mostly clean observations while ratios~$>1.3$ are a sign of severe soft-proton contamination.
In practice, we relied on EPIC count-rates calculated in the energy band $6-12$~keV for MOS and $5-7$ and $10-14$~keV for PN, but excluded the central area 10~arcmin of the FoV to minimise contamination by the sky signal at medium energies. In addition, since the cosmic-ray induced instrumental background also shows spatial variations, we computed the IN/OUT ratio for similar regions, flags and energies from our instrumental background template (Filter Wheel Closed, FWC, observations) and used it to renormalise our IN/OUT diagnostic. Experience shows that the criteria laid out by \citet{DeLuca2004} still apply with our modified version of the IN/OUT test.
We list the measured IN/OUT ratio of each observation in Table~\ref{tab:XMMobs}. These point to a low contamination by residual soft-proton emission and therefore we neglect this possible contribution in the spectral fitting (see Section~\ref{sub:spectralanalysis}).

The Solar Wind Charge Exchange \citep[SWCX,][]{Carter2008,Carter2011} may also be an important source of contamination in X-ray observations. We determine if our observations are affected by SWCX employing the method outlined in \citet{Carter2008} and \citet{Carter2011}. This procedure consists of comparing the count-rates of two light curves (with bin size of $1$~ks): the continuum-band ($2.5-5$~keV) and the line-band ($0.5-0.7$~keV). A scatter plot between the two bands is produced and a linear model fit is computed. If a high reduced $\chi^2$ value ($\geq1.2$) is obtained from the fit, then the observation is considered to have a SWCX enhancement. Moreover, the $\chi^2$ values for each individual light curve in terms of the deviation from their respective mean light curve are calculated. Then, the ratio between the $\chi^2$ value for the line band to the $\chi^2$ value for the continuum line is calculated. Values $\geq1.0$ of this ratio is expected in observations with high SWCX. None of the observations analysed in this work meets either of these two criteria, therefore we do not account for SWCX contamination in the spectral fitting analysis.

We have also checked for so-called `anomalous state' in the MOS detectors \citep{Kuntz2008}. Using the $0.2-0.9$~keV energy band, we found that CCD no. 5 of the MOS2 detector in the ZwCl1665 observation has a strong enhancement of photons at such energies. This affected CCD is excluded from further processing.

We also obtain a list of out-of-time events, which are added to the instrumental background (see Section~\ref{sub:spectralanalysis}) and then subtracted from the data set.

For the detection of point sources, we used a two-stage procedure. First, we applied the source detection method of \citet{Pacaud2006} in the $0.5-2$~keV energy band. Basically, we co-added images of the three EPIC detectors in this band, and the resulting image is filtered using the wavelet task {\tt MR\_FILTER} from the multiresolution package M{\tiny R/1} \citep{Starck1998}. The source detection and catalogue production are performed by the S{\tiny EXTRACTOR} software \citep{Bertin1996}. In a second step, we examined by eye the co-added images in the soft and hard band, $0.5-2$~keV and $2-10$~keV, respectively, to find possible undetected point sources. The corresponding regions of the final list of detected point sources were removed from the event lists in the subsequent analysis.

\begin{figure}[t]
\centering
\includegraphics[width=\columnwidth]{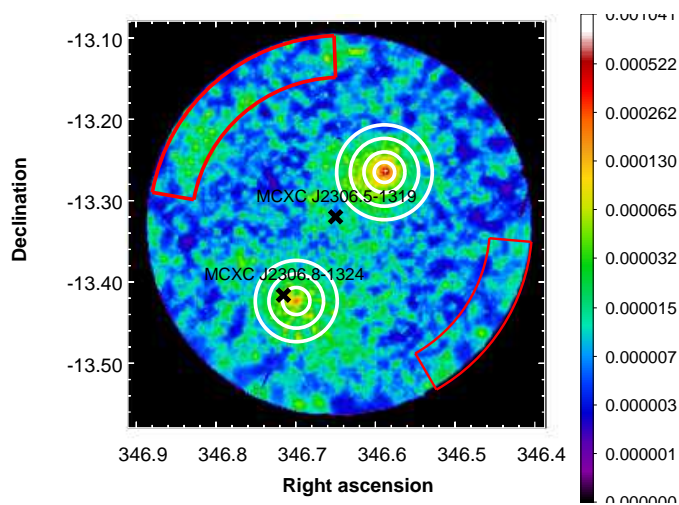}
\caption{\footnotesize{{\it XMM-Newton} smoothed, vignetting-corrected and background-subtracted image of the RXCJ2306.6-1319 observation in the $0.4-1.25$~keV energy band. The image is smoothed with a Gaussian kernel of sigma equal to three pixels. The concentric white circles show the chosen regions for the source spectral extraction. The regions delimited by the red curves are used to estimate the local sky background components. The units in the colour bar are count/s. The black crosses show the location of the MCXC clusters in the field. In this figure, North
is up and East to the left.}}
\label{fig:XMMSpecReg2306}
\end{figure}

\begin{figure}[t]
\centering
\includegraphics[width=\columnwidth]{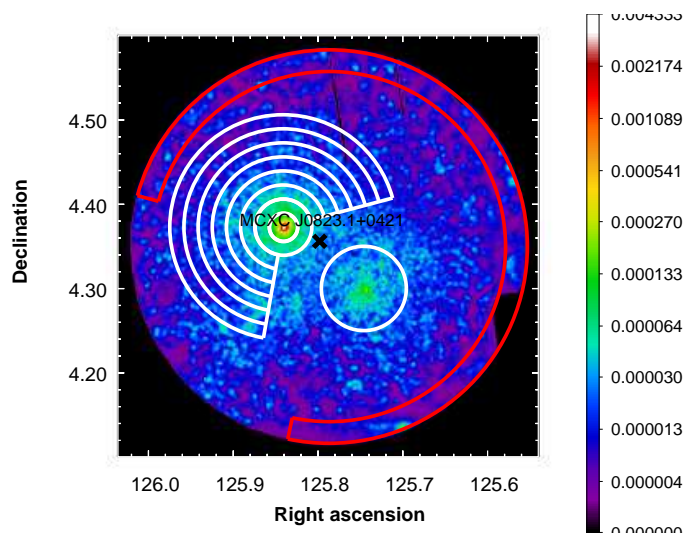}
\caption{\footnotesize{{\it XMM-Newton} smoothed, vignetting-corrected and background-subtracted image of the ZwCl1665 observation in the $0.4-1.25$~keV energy band. The caption information is the same as for Fig.~\ref{fig:XMMSpecReg2306}.}}
\label{fig:XMMSpecRegZwCl}
\end{figure}

\begin{figure}[t]
\centering
\includegraphics[width=\columnwidth]{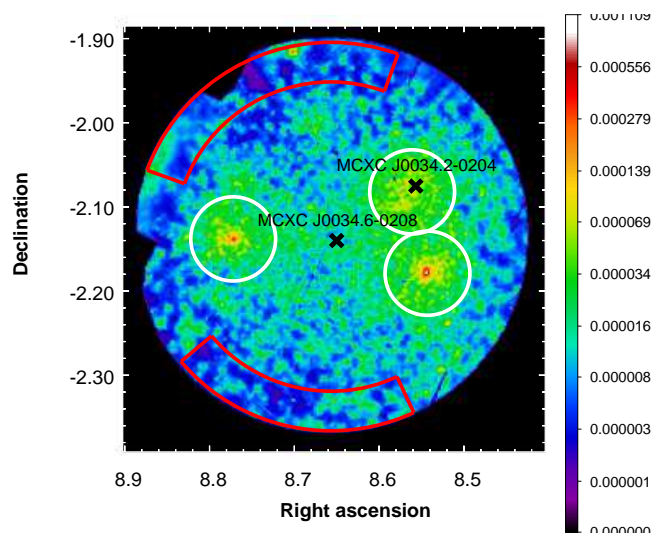}
\caption{\footnotesize{{\it XMM-Newton} smoothed, vignetting-corrected and background-subtracted image of the RXCJ0034.6-0208 observation in the $0.4-1.25$~keV energy band. The caption information is the same as for Fig.~\ref{fig:XMMSpecReg2306}.}}
\label{fig:XMMSpecReg0034}
\end{figure}


\subsection{X-ray spectral analysis}
\label{sub:spectralanalysis}

We performed a spectral analysis of the clusters in the $0.4-10$~keV energy band. The fitting procedure was performed using X{\tiny SPEC} v12.9.1h. In the following, we describe the modelling of the background and of the source.

\subsubsection{Background modelling}

It is well known that the modelling of the background is crucial to obtain reliable measurements of the properties of the ICM. The total background in X-ray observations consists of two main components: the non-vignetted particle background (PB) and the cosmic X-ray background (CXB). 

\begin{table*}[ht]
\centering
\caption{Extended structures identified in the different observations. Here, their X-ray peak coordinates, optical spectroscopic redshift, and X-ray spectroscopic redshift are displayed.}
\label{tab:grpscoorz}
\begin{threeparttable}
\renewcommand{\arraystretch}{1.35} 
  \begin{tabular}{l c c c c c c}
   \hline
   \hline
    Observation & Components & Identified & R.A. & Dec. & $z_\textrm{spec}^\textrm{opt}$ & $z_\textrm{spec}^\textrm{X-ray}$ \\
     & & Groups & (J2000) & (J2000) & & \\
   \hline
     RXCJ2306.6-1319 & SE & RXCJ2306.8-1324 & $346.699$ & $-13.424$ & $0.068\pm0.002$ & $0.065^{+0.011}_{-0.009}$ \\
      & NW & A2529 & $346.589$ & $-13.264$ & $0.109\pm0.003$ & $0.111^{+0.014}_{-0.011}$ \\
     ZwCl1665 & NE &ZwCl1665 & $125.841$ & $+4.372$ & $0.030\pm0.002$ & $0.028^{+0.002}_{-0.001}$\\
      & SW & SDSS-C4-DR3~1283 & $125.746$ & $+4.300$ & $0.097\pm0.003$ & $0.106^{+0.026}_{-0.032}$ \\
     RXCJ0034.6-0208 & E & -- & $8.583$ & $-2.070$ & $0.055$\tnote{$\dagger$} & $0.045^{+0.016}_{-0.012}$ \\
                     & NW & RXCJ0034.2-0204 & $8.772$ & $-2.138$ & $0.082$\tnote{$\dagger$} & $0.028^{+0.007}_{-0.005} $ \\
                     & SW & -- & $8.544$ & $-2.178$ & $0.057$\tnote{$\dagger$} & $0.082^{+0.007}_{-0.006}$ \\
   \hline
   \hline
  \end{tabular}
  \begin{tablenotes}\footnotesize
  \item[$\dagger$] These redshift values are taken from the closest brightest galaxy to their corresponding X-ray peak.
  \end{tablenotes}
\end{threeparttable}
\end{table*}

In {\it XMM-Newton} observations, the PB consists of a continuum component and fluorescence X-ray lines \citep[see][for details]{Snowden2008}. These lines affect the MOS and pn detectors differently: there are strong Al and Si lines in the $1.4-1.9$~keV energy band mainly affecting the MOS detectors, and other lines, e.g. Ni, Cu, and Zn, around the $7.4-9$~keV energy band showing in the pn detector. Since FWC observations are dominated by the instrumental background, we use them to subtract the internal instrumental background from our observations. For this, FWC observations must be re-normalized to match the level of PB present in the observations. Following the conclusions of \citet{Kuntz2008}, we renormalize the FWC data independently for each MOS CCD or PN quadrant, using their individual corners. For the MOS central CCDs, we compute the normalisation based on the corners of the surrounding CCDs that best correlate with them ([2,3,6,7] for MOS1 and [3,4,6] for MOS2). \citet{Zhang2009} have found that the $3-10$~keV energy band gives good re-normalization factors. We go a step further and narrow such energy band, avoiding the fluorescence X-ray lines present in the EPIC detectors. To calculate the FWC re-normalization factors we compute the count-rate, in the source and FWC observations, using photons out of the FoV for each CCD using the $2.5-5$ and $8-9$~keV energy bands for the MOS detectors and $2.5-5$~keV energy band for the pn detectors. We include Gaussian lines emission in the spectral fitting to account for residuals that exist after the PB subtraction.

The CXB is composed of a thermal emission component from the Local Hot Bubble (LHB), a second thermal emission component from the Galactic halo, and an extra-galactic component representing the unresolved emission from Active Galactic Nuclei (AGNs). Given that these distinct components have different spectral features, this background is more difficult to subtract from the observations. Instead, we model and fit the CXB components when doing the spectral analysis of our sources. We follow a similar spectral fitting methodology as the one described in \citet{Snowden2008}. The CXB is well described by a model that includes: an absorbed $\sim0.2$~keV thermal component representing the Galactic Halo emission \citep{McCammon2002}, an unabsorbed $\sim0.1$~keV thermal component representing the emission of the LHB, and an absorbed power-law with a fixed slope of $1.46$ representing the emission from the unresolved AGNs \citep{DeLuca2004}. In X{\tiny SPEC}, the thermal emission of the Galactic Halo and the LHB are modelled by the optically thin plasma model {\tt apec} \citep{Smith2001} with fixed solar abundances at $z=0$. We used the \citet{Asplund2009} abundance table for the relative abundance of heavy elements. The absorption is given by the Galactic column density of hydrogen, $n_\textrm{H}$, along the line-of-sight of each cluster, and it is modelled in X{\tiny SPEC} through {\tt phabs}. We obtained the $n_\textrm{H}$ value using the method of \citet[][see Table~\ref{tab:XMMobs}]{Willingale2013}.

\subsubsection{Spectral fitting}
\label{subsec:spectral_fitting}

The cluster thermal emission is modelled with an absorbed {\tt apec} model in X{\tiny SPEC}. We re-grouped all {\it XMM-Newton} spectra in order to secure at least $25$ photon counts per bin, which is necessary when using the $\chi^2$ minimization method. We fitted the {\it XMM-Newton} spectra in the $0.4-10$~keV energy band.

The spectral fitting of the data is done in two steps. First, for the modelling of the CXB we fit jointly the RASS spectra extracted from an annular region ($1-2$~deg from the cluster position) with {\it XMM-Newton background} spectra, with null or minimal cluster residual emission. The RASS spectra are obtained using a modified version of the available tool\footnote{https://heasarc.gsfc.nasa.gov/cgi\nobreakdash-bin/Tools/xraybg/xraybg.pl} at the HEASARC webpage. The main adjustment we did on this tool, was to mask out remaining sources in the RASS spectra. The {\it XMM-Newton background} spectra are obtained from annular sectors, $12.5-14$~arcmin, centred in the observation pointing (see red curves in Figs.~\ref{fig:XMMSpecReg2306}, \ref{fig:XMMSpecRegZwCl} and \ref{fig:XMMSpecReg0034}). In this joint fitting, the normalizations of the CXB components are linked across the spectra. The temperature, abundance and normalization of the cluster residual emission in the {\it XMM-Newton background} spectra are only linked across the EPIC detectors. These parameters, together with the normalizations of the residual Gaussian lines, are the only ones we allow to vary in the fitting. We take into account the proper corrections for the observed solid angle in the different spectra.

In a second part of the spectral modelling, we include the spectra extracted from concentric annuli (see white curves in Figs.~\ref{fig:XMMSpecReg2306}, \ref{fig:XMMSpecRegZwCl} and \ref{fig:XMMSpecReg0034}) to the joint spectral fitting. The best model parameters of the first fit are taken in this second step as initial values. As before, the normalizations of the CXB components, the normalizations of the residual Gaussian lines, the temperatures, metallicities, normalizations, and when the data allows it, the redshift in each cluster spectra are left free.


\section{Results}

\subsection{Visual impression}
\label{subsect:VisualDatabase}

The {\it XMM-Newton} images of RXCJ2306.6-1319, RXCJ0034.6-0208 and ZwCl1665 reveal double or triple extended structures (see Figs.~\ref{fig:XMMSpecReg2306}, \ref{fig:XMMSpecRegZwCl} and \ref{fig:XMMSpecReg0034}) centred around these RASS cluster positions. Our main findings of this first exercise are:

\begin{itemize}
 \item {\bf RXCJ2306.6-1319}: the {\it XMM-Newton} observation reveals a double extended structure aligned in the northwest-southeast direction (see Fig.~\ref{fig:XMMSpecReg2306}). The RXCJ2306.6-1319 position is located $\gtrsim 5$~arcmin from the centres of both extended objects. The southeastern extended source is $\sim1$~arcmin away from RXCJ2306.8-1324. Using the VizieR database \citep{Ochsenbein2000} we found that one FSC \citep[Faint Source Catalogue,][]{Voges2000} source is located nearby the southeastern extended component.

 \item {\bf ZwCl1665}: the {\it XMM-Newton} observation shows a double extended structure aligned in the northeast-southwest direction (see Fig.~\ref{fig:XMMSpecRegZwCl}). The position of ZwCl1665 is $\sim 2.5$~arcmin from the northeastern objects, while the southwestern component is $\sim 5$~arcmin away. A BSC \citep[Bright Source Catalogue][]{Voges1999} source coincides with the position of the northeastern structure. Although there are other four FSC sources in the field, none of them seems to be associated with the emission of the southwestern structure.
 
 \item {\bf RXCJ0034.6-0208}: the {\it XMM-Newton} observation of this cluster shows three extended sources (see Fig.~\ref{fig:XMMSpecReg0034}). The reported position of RXCJ0034.6-0208 lies in the middle of the three extended sources: $\sim 7$~arcmin from the eastern and southwestern components and $\sim6$~arcmin from the northwestern structure. The position of RXCJ0034.2-0204 agrees with an AGN within the northwestern extended source. In this latter case, it would have been difficult to resolve the AGN with ROSAT data. In fact, only one BCS source is in the field, located very close to the southwestern component. The Second ROSAT all-sky survey (2RXS) source catalogue \citep[][]{Boller2016} contains more sources ($7$) in this field, three of them are located nearby ($\sim1-2$~arcmin) each of the extended components.
\end{itemize}

\mbox{}

\begin{figure}
\centering
\includegraphics[trim=0 0 0 10,clip,width=\columnwidth]{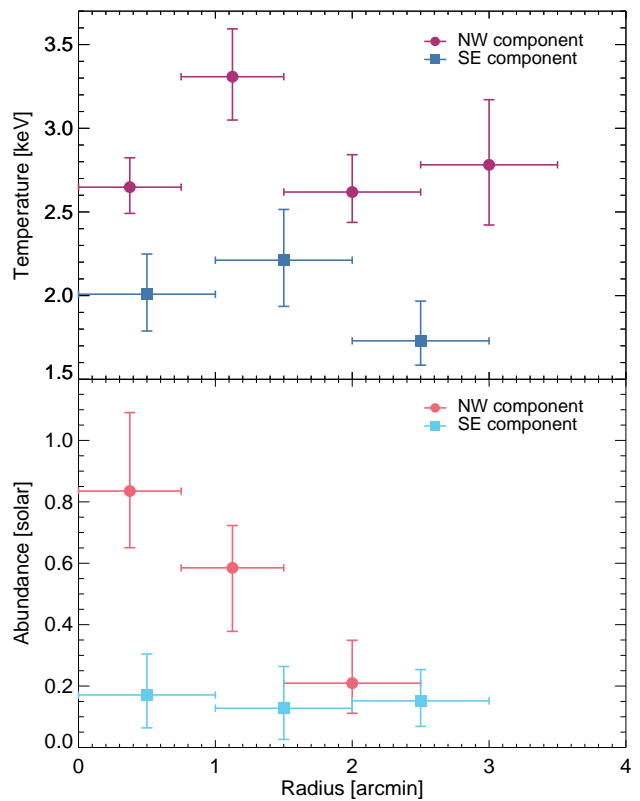}
\caption{\footnotesize{Temperature ({\it top panel}) and metal abundance ({\it bottom panel}) profiles for A2529 (northwestern structure in Fig.~\ref{fig:XMMSpecReg2306}, here in solid circles) and RXCJ2306.8-1324 (southeastern source in Fig.~\ref{fig:XMMSpecReg2306}, here in solid squares). The abundance in the outermost annulus of A2529 has been fixed to a value of $0.15$ (see text for details).}}
\label{fig:TAbprof2306}
\end{figure}

\begin{figure}
\centering
\includegraphics[trim=0 0 0 25,clip,width=\columnwidth]{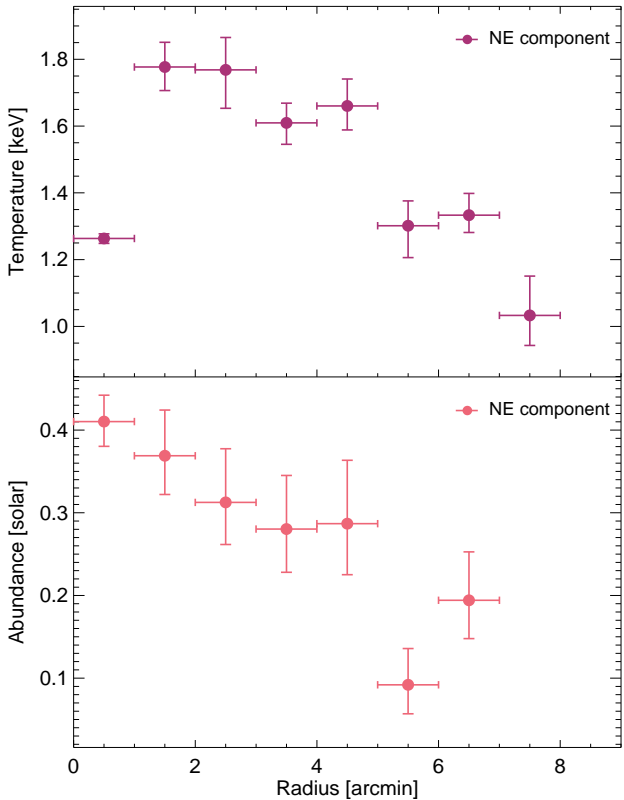}
\caption{\footnotesize{Temperature ({\it top panel}) and metal abundance ({\it bottom panel}) profiles of ZwCl1665. The measurements correspond to the white regions of the northeastern structure in Fig.~\ref{fig:XMMSpecRegZwCl}.}}
\label{fig:TAbprofZwCl}
\end{figure}

\subsection{Emission peak, X-ray redshift, temperature and abundance estimation of the cluster components}
\label{subsect:epzTZ}

We determined the X-ray emission peak of each extended structure in the observations using smoothed background-subtracted and vignetting-corrected images in the $0.4-1.25$~keV energy band. Table~\ref{tab:grpscoorz} shows the coordinates of the X-ray peak of each extended structure found in the observations.

We carried out an X-ray analysis for each component in each observation independently. We extracted spectra from concentric annuli of $1$~arcmin width (slightly smaller in the case of the northwestern structure in RXCJ2306.6-1319), or circular regions centred on the different X-ray structures as shown in Figs.~\ref{fig:XMMSpecReg2306}, \ref{fig:XMMSpecRegZwCl} and \ref{fig:XMMSpecReg0034} by the white regions. As described in Section~\ref{subsec:spectral_fitting}, we fitted the thermal cluster emission in each region with a thin plasma emission model {\tt apec} and we determined the temperature, metal abundance and redshift. Since the cluster morphology and data quality are different in each observation, we applied a slightly different spectral fitting approach to each cluster. In the following, we describe the details of the spectral fitting.

\begin{figure*}[t]
     \centering
     \includegraphics[trim=22 18 5 5,clip,width=0.65\textwidth]{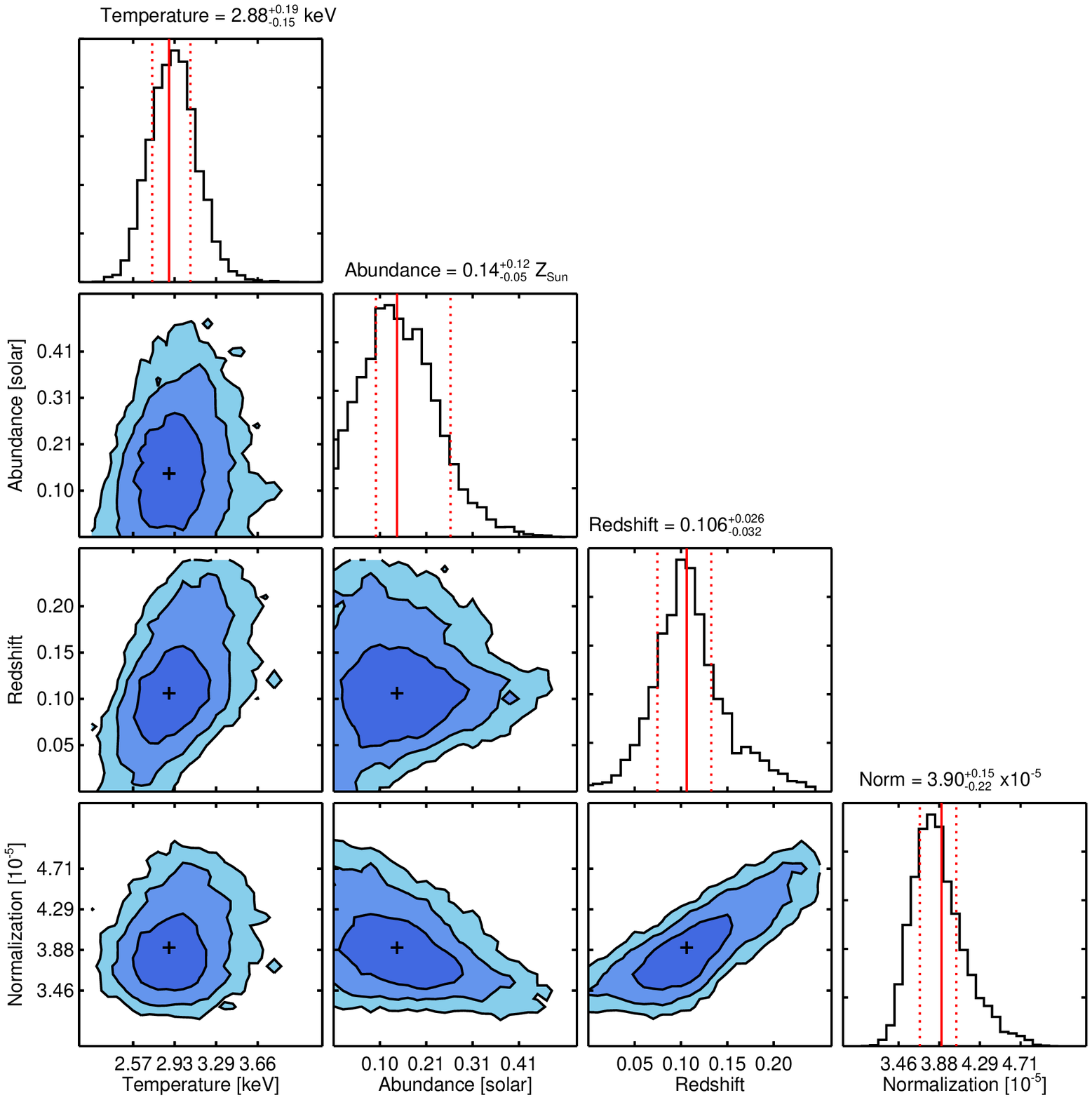}
     \caption{\footnotesize{Marginalized 2D and 1D constraints on the {\tt apec} model parameters obtained through an MCMC in X{\tiny SPEC} of the southwestern structure in the ZwCl1665 observation. The colours in the 2D plots represent the 68, 95 and 99\% credibility intervals. The red solid line on top of the 1D constraints and the cross overlaid on the 2D contours indicate the best-fitting X{\tiny SPEC} values, the red dotted lines show the 68\% confidence interval.}}
     \label{fig:TAbcorplotZwCl}
\end{figure*}

\begin{itemize}
 \item {\bf RXCJ2306.6-1319}: given the data quality of this observation, we were able to extract temperature and metal abundance profiles for both clusters for few annuli: $0-1,~1-2$ and $2-3$~arcmin for the southeastern structure, and $0-0.75,~0.75-1.50,~1.5-2.5$ and $2.5-3.5$~arcmin for the northwestern object (see Fig.~\ref{fig:TAbprof2306}). The abundance of the outer bin of the northwestern structure could not be constrained, therefore we fixed it to a value of $0.15~Z_\odot$, where $Z_\odot$ represents the solar abundance. The northwestern structure shows a temperature around $3$~keV, while the southeastern object has a temperature around $2$~keV. The temperature profile of the northwestern object exhibits a temperature decrement in the cluster centre, whereas the temperature profile of the southeastern component is consistent with a flat profile. The metal abundance profile of the northwestern structure decreases from $0.8~Z_\odot$ in the core to $0.2~Z_\odot$. In contrast, the metal abundance of the southeastern object remains flat at a value of $\sim0.2~Z_\odot$. In addition, we could free the redshift parameter during the spectral fitting of the northwestern structure, obtaining $z=0.111^{+0.014}_{-0.011}$. Figure~\ref{fig:histoZdist2306} shows the likelihood distribution of this redshift parameter\footnote{The likelihood is given by $\mathcal{L}=e^{-\chi^2/2}$, where $\chi^2$ is obtained from X{\tiny SPEC}.}. This redshift value is different from the value of $z=0.066$ reported in \cite{Piffaretti2011} for RXCJ2306.6-1319. The redshift in the southeastern structure could not be constrained in this joint spectral fitting, then the value of $z=0.066$, which corresponds to RXCJ2306.8-1324, was used \citep{Piffaretti2011}. If a circular region, of $3$~arcmin radius, is used for the spectral analysis of the southwestern component, we are able to constrain the redshift ($z=0.065^{+0.011}_{-0.009}$). The redshift difference between the two components is significant at $3\sigma$.

 \item {\bf ZwCl1665}: the data quality of this observation allows us to determine the temperature and metal abundance profiles up to $\sim8$~arcmin from the cluster centre of the most prominent extended source in the ZwCl1665 observation (see Fig.~\ref{fig:TAbprofZwCl}). As shown in Fig.~\ref{fig:XMMSpecRegZwCl}, the spectral extraction regions for the northeastern object avoid the overlapping region with the southwestern component. This cluster also shows a cool-core (CC) feature, i.e. the temperature in the innermost region has a lower value than its surroundings. The metal abundance profile exhibits a decreasing behaviour from $0.4~Z_\odot$ in the core down to $\sim0.1~Z_\odot$ in the outskirts. For this cluster, we also have enough data to free the redshift value during the spectral fitting. We obtain $z=0.028^{+0.002}_{-0.001}$, which is consistent with the redshift value of ZwCl1665 reported in \cite{Piffaretti2011}.

 The spectrum of the southwestern structure is heavily contaminated by the emission of the northeastern object, and this must be accounted for in the spectral fitting. In contrast to the previous spectral analysis, the cluster spectral emission of the southwestern structure (within $3$~arcmin radius of the X-ray peak, see Fig.~\ref{fig:XMMSpecRegZwCl}) is modelled by the background emission plus two {\tt apec} models, one representing the emission of the northeastern component and the other of the southwestern structure itself. The redshift of the {\tt apec} model of the northeastern object is kept fixed ($z=0.028$) since it has been well constrained and to reduce the number of free parameters during the fitting process. Moreover, we include the spectra of a region of the same area as the three last annuli of the northeastern component ($5-8$~arcmin, see Fig.~\ref{fig:XMMSpecRegZwCl}) to help to constrain the contaminating emission in the $3$~arcmin region of the southwestern source. We use the Markov-Chain Monte Carlo (MCMC) method with a Metropolis-Hastings algorithm in X{\tiny SPEC} in order to explore the full parameter space. The length of the MCMC chain is 200\,000 and we use flat priors. The contaminating emission of the northeastern component on the southwestern structure has a temperature of $kT=1.36^{+0.04}_{-0.03}$~keV and an abundance of $0.13^{+0.03}_{-0.04}~Z_\odot$, which are consistent with the values found for the three last individual annuli in the northeastern component (see Fig.~\ref{fig:TAbprofZwCl}). The emission of the southwestern structure seems to prefer a high redshift value of $0.106^{+0.026}_{-0.032}$, although this redshift value is consistent within $2.5\sigma$ with the one of the northeastern component ($z=0.028$). The results of the parameter sampling for the {\tt apec} model of the southwestern structure are shown in Fig.~\ref{fig:TAbcorplotZwCl}. Figure~\ref{fig:histoZdistZwCl1665} also shows the marginalized probability of the redshift. The temperature of the southwestern object is $kT=2.88^{+0.19}_{-0.15}$~keV and the abundance $0.14^{+0.12}_{-0.05}~Z_\odot$. When fixing the redshift of the southwestern structure to $z=0.03$ the abundances of the two {\tt apec} models could not be constrained. The southwestern structure is very faint in the available {\it Chandra} data. However, a spectral fitting seems to also favor a redshift $0.1-0.12$.
 
 \item {\bf RXCJ0034.6-0208}: the {\it XMM-Newton} observation of this cluster has a poor quality (see Sec.~\ref{sect:obsdataan}). Although the emission of the three extended structures in this observation might overlap (as in the case of ZwCl1665) and due to their low photon statistics, we carried out the X-ray analysis for each component independently, i.e. we assume no contaminating emission from the two other components while analysing the third one. We extracted spectra for each of the three extended structures within $3$~arcmin radius of the respective X-ray peak (see Fig.~\ref{fig:XMMSpecReg0034}). The joint MOS2 and pn spectral analysis determine the following temperatures, abundances and redshifts: for the eastern structure we found $kT=1.75^{+0.10}_{-0.16}$~keV, $Z=0.21^{+0.05}_{-0.04}~Z_{\odot}$, $z=0.045^{+0.016}_{-0.012}$, for the northwestern structure, $kT=2.45^{+0.09}_{-0.14}$~keV, $Z=0.50^{+0.06}_{-0.08}~Z_{\odot}$, $z=0.028^{+0.007}_{-0.005}$, and for the southwestern structure $kT=2.09^{+0.07}_{-0.07}$~keV, $Z=0.42^{+0.04}_{-0.06}~Z_{\odot}$, $z=0.082^{+0.007}_{-0.006}$. Only two of the three components agree within $2\sigma$ with the reported redshift values \citep[$z\sim0.08$,][]{Piffaretti2011} of RXCJ0034.6-0208 and RXCJ0034.2-0204 clusters. Given the short {\it XMM-Newton} observation it is difficult to detect the iron K and L lines of each component, which might lead to these differences in redshift. In Appendix~\ref{app:numberA} we explore further these differences by analysing each detector separately.

\begin{figure}
\centering
\includegraphics[trim=25 0 10 20,clip,width=\columnwidth]{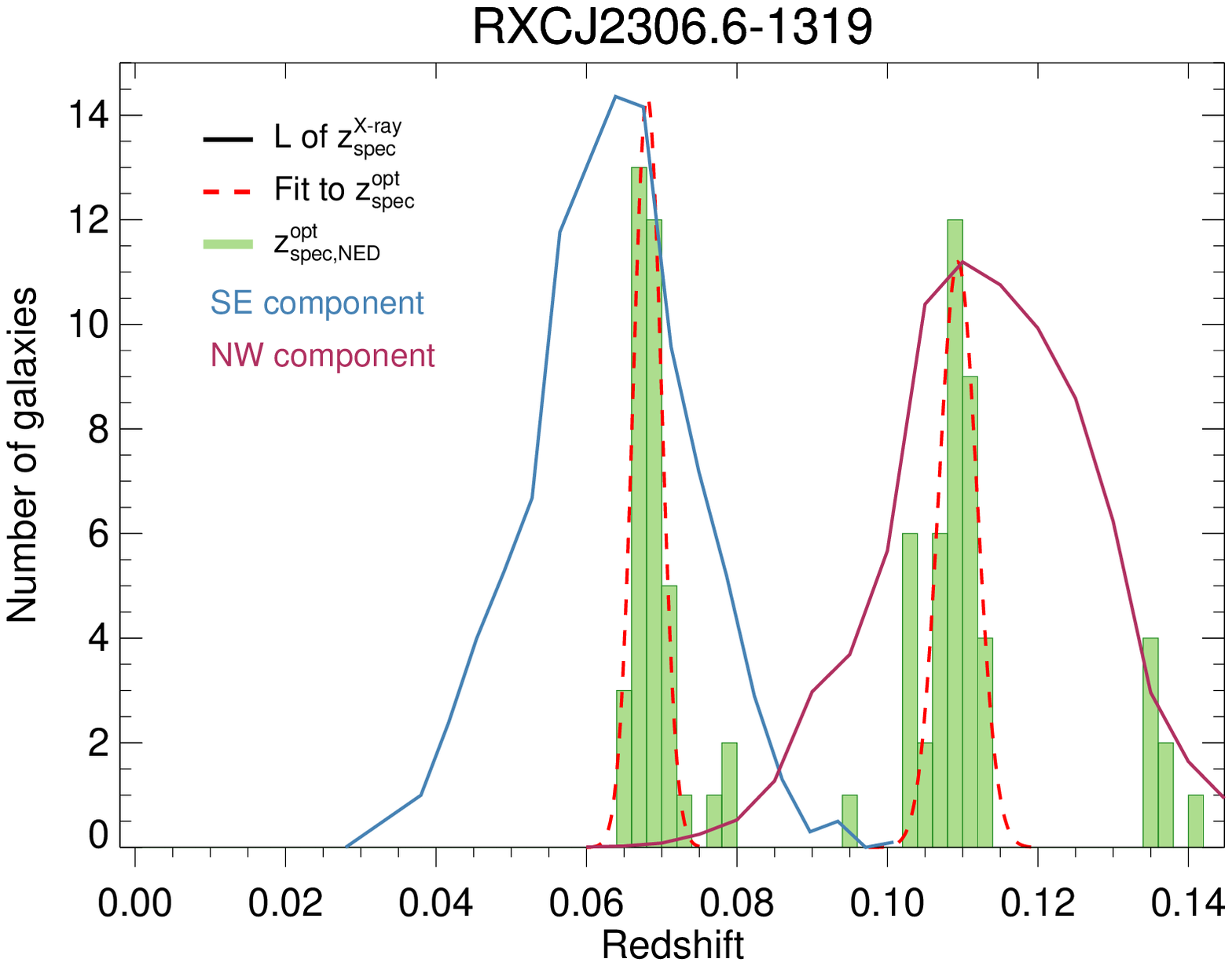}
\caption{\footnotesize Spectroscopic redshift histogram for galaxies within $30$~arcmin of the RXCJ2306.6-1319 cluster position. The peaks of this distribution are fitted with Gaussian functions (red dashed lines), obtaining $z=0.068\pm0.002$ for the low redshift peak, and $z=0.109\pm0.003$ for the high redshift structure. The solid lines shows the likelihood distribution of the redshift parameter obtained from the joint X-ray spectral fitting of the two structures shown in Fig.~\ref{fig:XMMSpecReg2306}. The blue line shows the southeastern component, while the purple one the northwestern object.}
\label{fig:histoZdist2306}
\end{figure}

\begin{figure}
\centering
\includegraphics[trim=25 0 10 20,clip,width=\columnwidth]{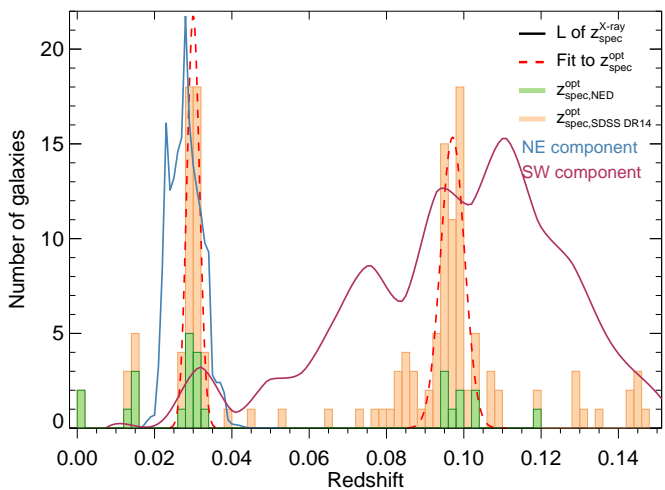}
\caption{\footnotesize{Spectroscopic redshift histogram for galaxies within $30$~arcmin of the ZwCl1665 cluster position. The caption information is the same as for Fig.~\ref{fig:histoZdist2306}, with the addition that the orange histogram shows the number of galaxies with available SDSS DR14 redshifts. The Gaussian fit of low redshift peak gives $0.030\pm0.002$, while for the high redshift one we obtain $0.097\pm0.003$. The likelihood distributions of the redshift parameter of the northeastern component is in purple, while the southwestern object is shown in blue.}}
\label{fig:histoZdistZwCl1665}
\end{figure}

\begin{figure}
\centering
\includegraphics[trim=25 0 10 20,clip,width=\columnwidth]{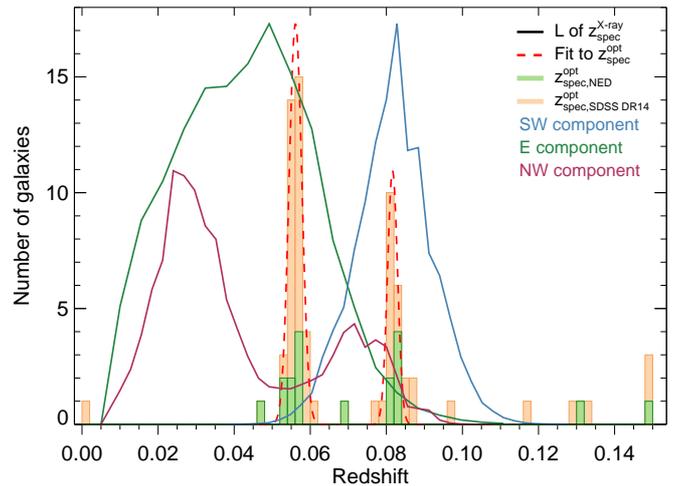}
\caption{\footnotesize{Spectroscopic redshift histogram for galaxies within $30$~arcmin of the RXCJ0034.6-0208 cluster position. The caption information is the same as for Fig.~\ref{fig:histoZdistZwCl1665}. The Gaussian fit of the low redshift peak gives $z=0.056\pm0.002$, while the high redshift structure peaks at $z=0.082\pm0.001$. The likelihood distributions are: in green the eastern component, in purple the northwestern object and in blue the southwestern structure.}}
\label{fig:histoZdist0034}
\end{figure} 
 
\end{itemize}

\subsection{Database search and optical redshift analysis}
\label{subsec:zoptspec}

We combined the available information from the NASA/IPAC Extragalactic Database\footnote{https://ned.ipac.caltech.edu/} (NED) and SDSS~DR14\footnote{http://www.sdss.org/dr14/} to construct spectroscopic redshift distributions of each observation (see Figs.~\ref{fig:histoZdist2306}, \ref{fig:histoZdistZwCl1665} and \ref{fig:histoZdist0034}). For this, we select all galaxies with spectroscopic redshift within $30$~arcmin of the cluster positions. All the individual redshift distributions of these observations reveal two structures (bi-modal distributions) at different redshifts. In these distributions, one of the peaks is consistent with the MCXC redshift (see Table~\ref{tab:knownproperties}).

We fit a Gaussian function to each of the peaks in the redshift distribution to obtain a redshift estimation of the different structures. For the structures in RXCJ2306.6-1319 and ZwCl1665, these values are displayed in Table~\ref{tab:grpscoorz} under the column labeled $z_\textrm{spec}^\textrm{opt}$, and they are consistent (within $1\sigma$) with the one obtained from our X-ray analysis (column $z_\textrm{spec}^\textrm{X-ray}$ also in Table~\ref{tab:grpscoorz}).

We further investigate these extended structures by searching in NED for known galaxy groups and/or clusters with available spectroscopic redshift information within $15$~arcmin of each observation. Our findings are summarized as following:
\begin{itemize}
 \item {\bf RXCJ2306.6-1319}: the northwestern object is located nearby A2529 ($<1$~arcmin). The redshift of this cluster is $z=0.111$ \citep{Struble1999}, which is in agreement with the redshift obtained from our X-ray spectroscopic analysis.
 \item {\bf ZwCl1665}: within $\sim6$~arcmin of the northeastern object position, NED shows several galaxy groups and clusters with spectroscopic redshift of values $z\sim0.03$, which agree with the value obtained from our X-ray spectroscopic analysis. One of these groups is SDSS-C4-DR3~1356 \citep[$z=0.03$ from 34 galaxies,][]{VonDerLinden2007}. The location of the southwestern structure agrees ($<0.1$~arcmin) with the SDSS-C4-DR3~1283 cluster position \citep[$z=0.095$ from 23 galaxies,][]{VonDerLinden2007}. This latter extended substructure is also located $\sim1$~arcmin away from a cluster in the recently published galaxy cluster catalogue by \citet{Banerjee2018} (ID 10959 with $z=0.0958$). This redshift is also in agreement with the redshift obtained from our X-ray analysis.
 \item {\bf RXCJ0034.6-0208}: NED does not show other galaxy groups or clusters besides RXCJ0034.6-0208 and RXCJ0034.2-0204. The X-ray peak positions of the eastern and southwestern extended components coincide with bright elliptical galaxies (see Fig.~\ref{fig:Zdist0034}) with $z\sim0.055$ (from SDSS DR14). On the contrary, the X-ray peak of the northwestern component is located nearby an elliptical galaxy with a spectroscopic redshift value of $z=0.082$. The redshifts of these bright elliptical galaxies are displayed in Table~\ref{tab:grpscoorz}. Only the redshift of the brightest galaxy in the eastern structure is in agreement within $1\sigma$ with the redshift obtained our X-ray analysis ($z\sim0.045$).
\end{itemize}

\begin{figure}[t]
\centering
\includegraphics[width=\columnwidth,trim={2.5cm 0.5cm 4.0cm 1.8cm},clip]{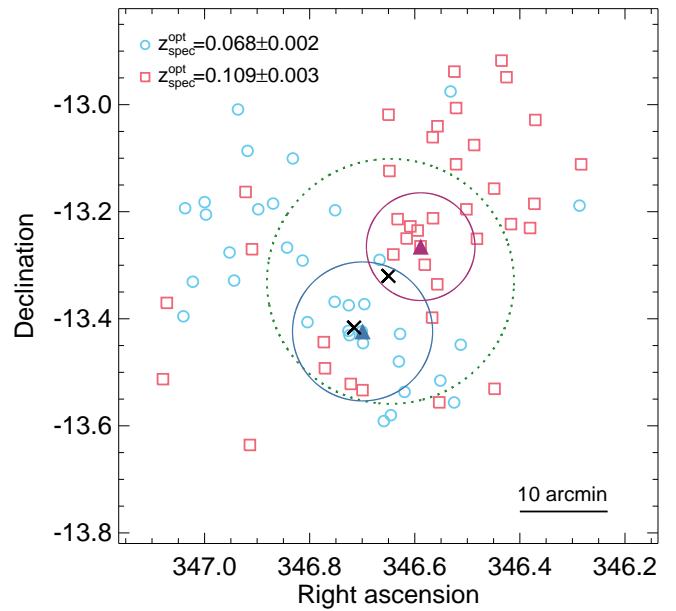}
\caption{\footnotesize{Projected redshift distribution of galaxies with spectroscopic redshift information for the RXCJ2306.6-1319 observation. Light blue circles correspond to galaxies with redshift values of $\sim0.068$, and pink squares of $\sim0.109$. The black crosses correspond to the position of the MCXC clusters in the observation and the green dotted circle shows the FoV of {\it XMM-Newton} (see Fig.~\ref{fig:XMMSpecReg2306} for comparison). The purple filled triangle marks the X-ray peak position of A2529, while the blue one of RXCJ2306.8-1324. Their corresponding circles show $r_{500}$}. In this figure, North is up and East to the left.}
\label{fig:Zdist2306}
\end{figure}

\begin{figure}[t]
\centering
\includegraphics[width=\columnwidth,trim={2.5cm 0.5cm 4cm 1.8cm},clip]{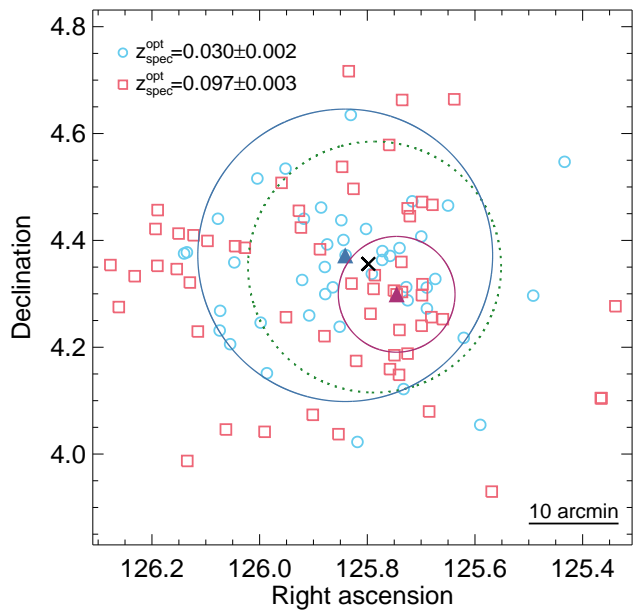}
\caption{\footnotesize{Projected redshift distribution of galaxies with spectroscopic redshift information for the ZwCl1665 observation. The caption information is the same as for Fig.~\ref{fig:Zdist2306}, with the exception that light blue circles correspond to galaxies with redshift values of $\sim0.030$ and pink squares of $\sim0.097$. The purple filled triangle shows the X-ray peak position of ZwCl1665, while the blue one that of SDSS-C4-DR3~1283.}}
\label{fig:ZdistZwCl1665}
\end{figure}

\begin{figure}
\centering
\includegraphics[width=\columnwidth,trim={2.5cm 0.5cm 4cm 1.8cm},clip]{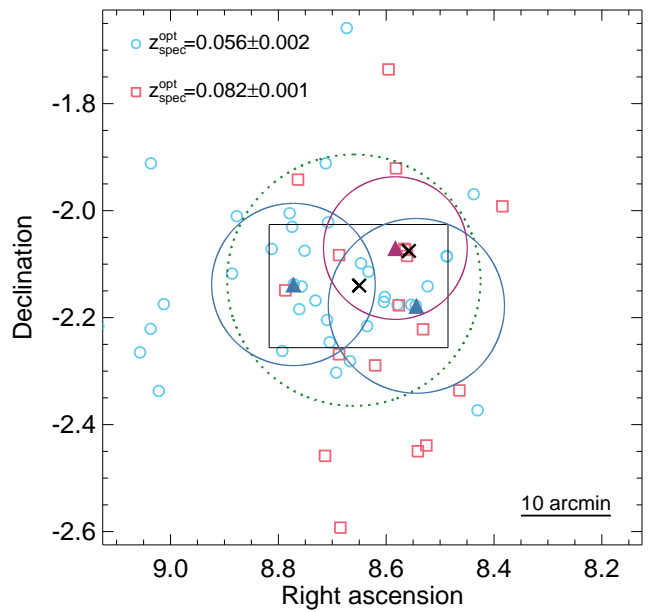}
\includegraphics[width=0.99\columnwidth]{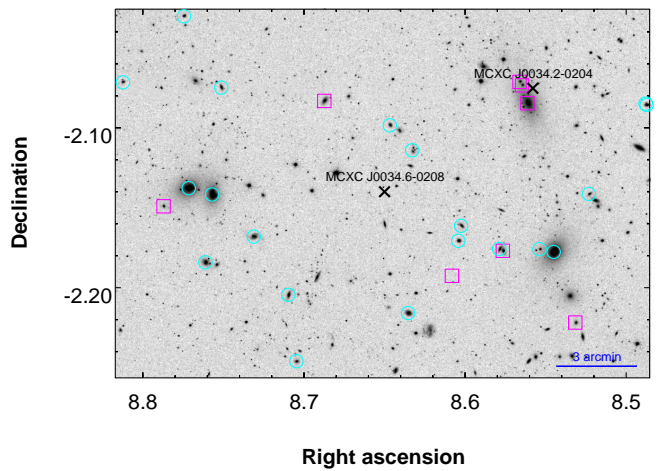}
\caption{\footnotesize{{\it Top}: Projected redshift distribution of galaxies with spectroscopic redshift information for the RXCJ0034.6-0208 observation. The caption information is the same as for Fig.~\ref{fig:Zdist2306}, with the exception that light blue circles correspond to galaxies with redshift values of $\sim0.056$ and pink squares of $\sim0.082$. The filled triangles shows the X-ray peak positions of the three extended substructures in the field. {\it Bottom}: SDSS {\it r}~-band image of the region delimited by the black-solid line in the top panel.}}
\label{fig:Zdist0034}
\end{figure}

We applied a $3\sigma$ clipping to each redshift structure in their distributions to remove outliers, and display the projected distribution of the galaxies (see Figs.~\ref{fig:Zdist2306}, \ref{fig:ZdistZwCl1665} and \ref{fig:Zdist0034}). Although the redshift information is not complete, it can give us a good idea of the galaxy redshift distribution in the plane of the sky. Figure~\ref{fig:Zdist2306} shows a projected redshift separation in the RXCJ2306.6-1319 region, with a higher concentration of galaxies with $z\sim0.109$ in the northwest direction. The case of ZwCl1665 is not as clear as RXCJ2306.6-1319, galaxies of the two redshifts estimations cover the whole plane of the sky, however, there is a slight over-density of galaxies with $z=0.097\pm0.003$ close to the position of SDSS-C4-DR3~1283. Around the eastern and southwestern components of RXCJ0034.6-0208 we can see a higher concentration of galaxies around $z\sim0.056$, while the northwestern structure lies close by to galaxies with $z\sim0.082$.

Given the results of the optical spectroscopic redshift and X-ray spectral analysis, we relate the southeastern and northwestern structures in the observation RXCJ2306.6-1319, to RXCJ2306.8-1324 and A2529, respectively. In the same way, we connect the southwestern extended component in ZwCl1665 to cluster SDSS-C4-DR3~1283. The case of RXCJ0034.6-0208 is more complex, given that the X-ray spectral analysis seems to give contradicting redshift information in comparison with the optical data.

\subsection{Global properties}

\subsubsection{Mass and radius}

In this section we derive the global properties of each galaxy cluster namely the emission weighted global temperature, $kT$, the global metallicity, $Z$, the mass, $M_{500}$, and the corresponding radius, $r_{500}$, flux, $F_{500}$, and luminosity, $L_\textrm{500}$.

For some galaxy groups and clusters the gas temperature, $kT$, decreases in the central regions, resulting in CC systems. In order to avoid the effect of CC, i.e. biasing the global gas temperature estimate low, for clusters with enough photon counts and with good data quality we measure the global gas temperature within $0.1-0.5r_{500}$ \citep{Zhang2006}, for the other clusters the temperature is determined within $0.5r_{500}$ (see Table~\ref{tab:grpsderprop}). To estimate $M_{500}$ we use the scaling relation obtained by \citet{Lovisari2015}:
\begin{equation}
    \log_{10}\Bigg(\frac{M_{500}}{5\times10^{13}~\textrm{M}_\odot}\Bigg)=1.65\log_{10}\Bigg(\frac{kT}{2~\textrm{keV}}\Bigg)+0.19.
\end{equation}
We choose this scaling relation since it has been determined using a galaxy group sub-sample of eeHIFLUGCS with similar temperatures as the ones we find in this work. An initial temperature is obtained by performing a spectral fitting to the data in a given aperture. The global gas temperature and $M_{500}$ are evaluated iteratively until we obtain a stable temperature. We determine $r_{500}$ from $M_{500}$ using
\begin{equation}
    r_{500}=\Bigg(\frac{M_{500}}{\frac{4\pi}{3}500\rho_\textrm{cr}(z)}\Bigg)^{\frac{1}{3}},
\end{equation}
where $\rho_\textrm{cr}(z)$ is the critical density at the cluster redshift. The estimated $kT$, $Z$, $M_{500}$ and $r_{500}$ of each galaxy cluster studied in this work are listed in Table~\ref{tab:dtrmndprprts}. We notice that four of the analysed groups in this work have abundance values $Z\lesssim0.2~Z_{\odot}$, either within $0.5r_{500}$ or $0.1-0.5r_{500}$. In a study of $\sim200$ galaxy groups and clusters, \citet{Lovisari2019} found such low metallicity for only $1-2$ objects within $0.3r_{500}$.

\begin{table*}[ht]
\centering
\caption{Derived global properties for the galaxy groups. The flux, $f_{500}$, and luminosity, $L_{500}$, are obtained in the $0.1-2.4$~keV observer and rest-frame energy band, respectively.}
\label{tab:grpsderprop}
\begin{threeparttable}
\renewcommand{\arraystretch}{1.35} 
 \centering
  \begin{tabular}{l c c c c c c}
   \hline
   \hline
    Cluster & $kT$ & $Z$ & $r_{500}$ & $M_{500}$ & $f_{500}$ & $L_{500}$\\
     & [keV] & [$Z_{\odot}$] & [kpc] & [$10^{13}$~M$_\odot$] & [$10^{-12}$~erg~s$^{-1}$~cm$^{-2}$] & [$10^{43}$~erg~s$^{-1}$] \\
   \hline
     RXCJ2306.8-1324~\tnote{$\dagger$} & $1.88^{+0.16}_{-0.16}$ & $0.21^{+0.09}_{-0.07}$ & $625^{+42}_{-18}$ & $6.98^{+1.07}_{-0.98}$ & $1.05\pm0.12$ & $1.17\pm0.13$ \\ 
     A2529~\tnote{$\dagger$} & $2.70^{+0.18}_{-0.14}$ & $0.38^{+0.08}_{-0.11}$ & $752^{+41}_{-13}$ & $12.73^{+1.45}_{-1.32}$& $1.54\pm0.09$ & $4.52\pm0.27$\\ 
     ZwCl1665~\tnote{$\dagger$} & $1.52_{-0.04}^{+0.04}$ & $0.20^{+0.02}_{-0.02}$ & $554^{+13}_{-11}$ & $4.93^{+0.34}_{-0.31}$ & $5.68\pm0.44$  & $1.17\pm0.09$\\ 
     SDSS-C4-DR3~1283~\tnote{*} & $2.84^{+0.18}_{-0.16}$ & $0.16^{+0.08}_{-0.07}$ & $757^{+27}_{-29}$ & $13.80^{+1.61}_{-1.45}$ & $1.73\pm0.10$ & $3.97\pm0.24$ \\ 
     RXCJ0034.6-0208 E~\tnote{*} & $1.70^{+0.09}_{-0.06}$ & $0.16^{+0.03}_{-0.03}$ & $582^{+16}_{-17}$ & $5.92^{+0.53}_{-0.50}$ & $3.22\pm0.25$ & $2.29\pm0.18$\\ 
     RXCJ0034.6-0208 NW~\tnote{*} & $2.68^{+0.11}_{-0.11}$ & $0.52^{+0.10}_{-0.09}$ & $742^{+21}_{-20}$ & $12.55^{+1.10}_{-1.01}$ & $4.24\pm0.35$ & $6.87\pm0.57$\\ 
     RXCJ0034.6-0208 SW~\tnote{*} & $2.07^{+0.07}_{-0.07}$ & $0.30^{+0.05}_{-0.04}$ & $648^{+16}_{-15}$ & $8.17^{+0.62}_{-0.56}$ & $4.27\pm0.44$ & $3.26\pm0.33$\\ 
   \hline
   \hline
  \end{tabular}
    \begin{tablenotes}\footnotesize
  \item[$\dagger$] Global temperature determined within $0.1-0.5r_{500}$.
  \item[*] Global temperature determined within $0.5r_{500}$.
  \end{tablenotes}
\end{threeparttable}
\label{tab:dtrmndprprts}
\end{table*}

\subsubsection{Flux and luminosity}

Given that most of the newly identified sub-clusters overlap in the plane of the sky, we needed to fit their surface brightness profiles simultaneously to estimate their individual fluxes. We chose the $0.4-1.25$ keV energy band for this, as the cluster signal vastly dominates over the background in this soft band, while the absence of strong fluorescence lines in the instrumental background ensure minimal systematic errors - Al K$\alpha$ and Si K$\alpha$ at $\sim 1.5$ and $\sim 1.7$~keV  respectively prevented us form using higher energies.  Each of the two or three sub-clusters are modelled as a set of concentric 3~arcsec width annuli whose $0.4-1.25$~keV surface brightnesses are simultaneously fitted. The position, instrument-dependent {\it XMM-Newton} PSF is taken into account during the fit. For a set of parameters, we used the best fit spectral model from the results discussed in Section~\ref{subsect:epzTZ} to convert the flux profile into count-rates for each instrument. We combined these with a sky background count-rate also derived from our best fit spectral model, the exposure map and template-based instrumental background maps to form a counts map. Poisson statistics over the whole images is then used to derive the best-fit surface brightness profiles. Statistical uncertainties are finally determined from using a Metropolis Markov-Chain Monte-Carlo analysis, based on the covariance of 100\,000 element chains.

We determine the flux in $r_{500}$ for each cluster by integration of the surface brightness profiles within the radii reported in Table~\ref{tab:grpsderprop} (all components can be traced out to $r_{500}$). The statistical uncertainties are combined in quadrature with the errors on the background flux in the same aperture obtained from our spectral modelling. Finally, the cluster optical redshift and best fit model is used to convert this flux into a luminosity, ignoring the uncertainty on the cluster spectral model in the process.  

The measured $f_{500}$ and $L_{500}$ in the $0.1-2.4$~keV energy band, respectively, for each cluster are displayed in Table~\ref{tab:dtrmndprprts}.

\begin{figure}[t]
\centering
\includegraphics[width=\columnwidth,trim={2.0cm 0.0cm 3.2cm 1.0cm},clip]{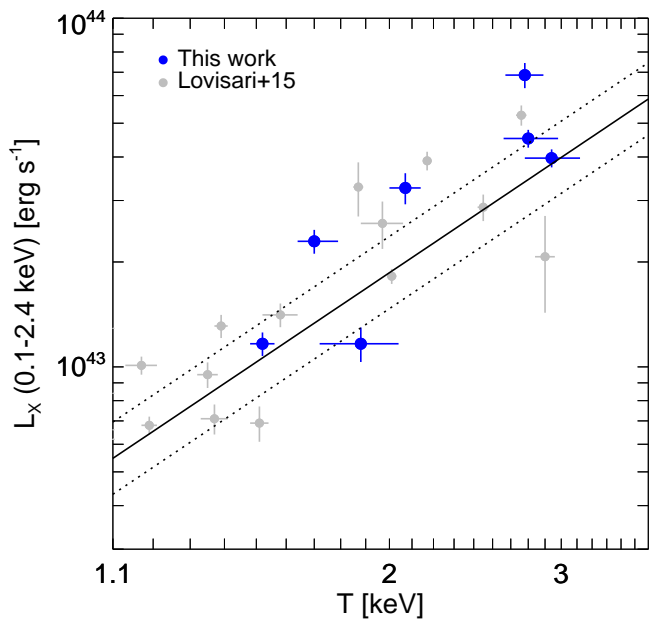}
\caption{\footnotesize{Comparison of the global properties, luminosity and temperature, of the analysed clusters in this work (blue circles) with the group sample studied in \cite{Lovisari2015} (grey points). The solid/dashed blue lines correspond to the non-bias corrected $L_\textrm{X}-T$ scaling relation, from the same authors, and its total scatter.}}
\label{fig:ltrel}
\end{figure}

Our flux measurements show that out of all the individual systems studied in this work only the northeastern component in the ZwCl11665 observation has a flux value high enough ($5.7\times10^{-12}$~erg~s$^{-1}$~cm$^{-2}$) to be kept in eeHIFLUGCS (see Sect.~\ref{sect:knowninfo}).

Figure~\ref{fig:ltrel} compares our luminosity and temperature measurements with the ones of \cite{Lovisari2015} group sample. We chose this sample because \cite{Lovisari2015} obtained its measurements using {\it XMM-Newton} observations as well, and the procedures for determining each observable are similar to the ones used in this work. The corresponding non-bias corrected relation from that work is also shown in Fig.~\ref{fig:ltrel}, together with its total scatter on the luminosity ($\sim20\%$). Taking into account the total scatter, the agreement is good despite that the group sample of \cite{Lovisari2015} covers only lower redshift objects ($z<0.035$). 

\begin{table}[t]
\caption{Comparison of fluxes in the $0.1-2.4$~keV energy band (see text for details).}
 \centering
\begin{threeparttable}
\renewcommand{\arraystretch}{1.2} 
  \begin{tabular}{l c c c}
   \hline
   \hline
    Cluster & $f_{500}^\textrm{MCXC}$ & $f_{500}^\textrm{GCA}$ & $f_{500}^{XMM}$ \\
     & \multicolumn{3}{c}{[$10^{-12}$~erg~s$^{-1}$~cm$^{-2}$]} \\
   \hline
     RXCJ2306.6-1319 & $5.58$ & $5.55\pm1.00 $ & $2.59\pm0.15$ \\
     ZwCl1665 & $9.81$ & $5.80\pm1.10$ & $7.40\pm0.45$ \\
     RXCJ0034.6-0208 & $7.95$ & $8.24\pm0.71$ & $11.7\pm0.62$ \\
   \hline
   \hline
  \end{tabular}
\label{tab:fluxcomparison}
\end{threeparttable}
\end{table}

Finally, we compare the {\it XMM-Newton} determined fluxes for the different cluster systems with the overall flux estimation from ROSAT data. With this, we aim to examine the reliability of the the previous flux measurements. For this, we compare three quantities: the flux expected from the MCXC catalogue using the redshift information from MCXC, the ROSAT flux estimation using the growth curve analysis (GCA) and the {\it total} flux from {\it XMM-Newton} using $z_\textrm{spec}^\textrm{opt}$ in Table~\ref{tab:grpscoorz}. The MCXC fluxes, $f_{500}^\textrm{MCXC}$, are estimated using the luminosities and redshifts from MCXC, together with the luminosity-temperature, $L_\textrm{X}-T$ scaling relation of \cite{Reichert2011}. To relate observer frame and rest frame quantities, we apply a $k$-correction assuming an abundance of $0.3~Z_\odot$ \citep{Asplund2009}. The CGA fluxes, $f_{500}^\textrm{GCA}$, are obtained using a similar photometric analysis like the one described in \cite{Boehringer2000,Boehringer2001}, and it is fully described in \cite{Xu2018}. In our case, this method integrates the background subtracted count-rate in the ROSAT hard band ($0.5-2$~keV) up-to $r_{500}$. The {\it total} flux from {\it XMM-Newton}, $f_{500}^{XMM}$, is obtained by adding the individual fluxes (Table~\ref{tab:dtrmndprprts}) of the corresponding components in the cluster systems. Table~\ref{tab:fluxcomparison} shows the values of the different estimated fluxes. There is a good agreement between the $f_{500}^\textrm{MCXC}$ and $f_{500}^\textrm{GCA}$ measurements for the RXCJ2306.6-1319 and RXCJ0034.6-0208 systems, which indicates that our method is consistent with the one in \cite{Boehringer2000,Boehringer2001}. However, the $f_{500}^{XMM}$ values of these systems differs from those measurements. This contrasts with the good agreement between the $L_{500}$ values from MCXC (see Table~\ref{tab:knownproperties}) and the {\it total} {\it XMM-Newton} luminosity, i.e. when adding the individual {\it XMM-Newton} luminosities we obtain, $\sim5.7$ and $\sim12.4\times10^{43}$~erg~s$^{-1}$ for RXCJ2306.6-1319 and RXCJ0034.6-0208, respectively (see Table~\ref{tab:dtrmndprprts}). The difference between ROSAT and {\it XMM-Newton} fluxes could be due to different redshift and modelling assumptions. We mainly attribute the disagreement between $f_{500}^\textrm{GCA}$ and $f_{500}^{XMM}$ fluxes for these two systems to the different redshift and modelling assumptions used to characterize each individual object. While for $f_{500}^\textrm{GCA}$ the MCXC redshift is used, the $f_{500}^{XMM}$ is obtained from systems at different redshifts. Moreover, the flux discrepancy can be partially justified in system RXCJ2306.6-1319 by better detection of point-sources in the {\it XMM-Newton} data. But this does not explain that $f_{500}^{XMM}$ is larger than $f_{500}^\textrm{GCA}$ in the other two systems.

Within $1.5\sigma$ the $f_{500}^\textrm{GCA}$ and $f_{500}^{XMM}$ values for ZwCl1665 agree, however the value of $f_{500}^\textrm{MCXC}$ is higher than those values. The parent catalogue of ZwCl1665 is BCS from \cite{Ebeling1998}, who uses a Voronoi Tessellation and Percolation (VTP) method to estimate the count-rate and extent of the clusters in their catalogue. For this particular cluster, they obtain a higher count-rate in a smaller extraction radius. These differences with respect to the GCA method may explain the factor of $\sim2$ between the $f_{500}^\textrm{GCA}$ and $f_{500}^\textrm{MCXC}$ for ZwCl1665.


\section{Discussion and conclusion}

The {\it XMM-Newton} and {\it Chandra} follow-up of the eeHIFLUGCS sample of galaxy clusters has revealed that three clusters are the result of projection effects of two (or three) physically independent clusters at different redshifts. Our X-ray spectral analysis reveals that the redshift of at least one of the components in each of these multiple systems has a different redshift from the one reported in the literature \citep{Piffaretti2011}. The measured X-ray redshifts are confirmed by a spectroscopic optical analysis, which shows bi-modal redshift distributions in the sky region of the analysed clusters. We report on the global properties of each component: the gas temperature, the total masses within $r_{500}$, the flux and luminosity.

In the RXCJ2306.6-1319 observation, we found two clusters at redshifts $z\sim0.06$ and $z\sim0.11$. The cluster at lower redshift is consistent in position and redshift with the RXCJ2306.8-1324 cluster reported by \cite{Cruddace2002} in the SGP catalogue. Our analysis reveals that the higher redshift cluster is in location and redshift agreement with A2529. RXCJ2306.8-1324 is less massive ($M_{500}\sim7\times10^{13}$~M$_\odot$) than A2529 ($M_{500}\gtrsim1.2\times10^{14}$~M$_\odot$). Our surface brightness analysis shows that both clusters have fluxes below the eeHIFLUGCS flux limit ($5\times10^{-12}$~erg~s$^{-1}$~cm$^{-2}$), therefore both clusters are no longer part of this cluster sample. Moreover, RXCJ2306.6-1319 appears in the supercluster sample of \cite{Chon2013}, with a multiplicity of $2$, i.e. the number of clusters that form the supercluster, at $z=0.067$. This supercluster sample is based on the REFLEX II catalogue \citep{Boehringer2013}, which is an extension of REFLEX with a lower flux limit. If indeed, \cite{Boehringer2013} included RXCJ2306.8-1324 in REFLEX II (the REFLEX II is not publicly available), then this cluster together with RXCJ2306.6-1319 would look as a supercluster. However, our findings show that RXCJ2306.6-1319 is indeed a single cluster at higher redshift.

The X-ray spectral analysis of the two extended components in the ZwCl1665 observation also shows that both structures are located at different redshifts ($z\sim0.03$ and $z\sim0.10$). ZwCl1665 is the lower redshift galaxy group, and it is confirmed by several optical counterparts. ZwCl1665 is a cool system ($kT\sim1.5$~keV) with a mass of $M_{500}\sim5\times10^{13}$~M$_\odot$. The second group identified in the ZwCl1665 observation has a lower flux due to its higher redshift but is hotter ($\sim3$~keV), and more massive ($M_{500}\gtrsim1.3\times10^{14}$~M$_\odot$) than ZwCl1665. We cross-identified this system with the SDSS-C4-DR3~1283 cluster \citep[][]{VonDerLinden2007}. The new estimations of the flux of ZwCl1665 still locate it slightly above the eeHIFLUGCS flux limit.

Our joint X-ray and optical analysis of the three components in observation RXCJ0034.6-0208 indicates that two of such structures are located at similar redshift ($z\sim0.05$), while the third one is at a higher redshift ($z\sim0.08$). The X-ray peaks of the three components nicely coincide with giant elliptical galaxies located at different redshifts, and the galaxy redshift distribution is bi-modal. However, {\it XMM-Newton} presented some anomalies while observing this target: there is no data for the MOS1 camera and the X-ray spectral results of two of the three components differ between the MOS2 and pn cameras. The poor quality of the X-ray observation does not allow us to perform a deeper and joint analysis of this system with the available {\it Planck} data, as in \cite{Planck2013}. With a deeper {\it XMM-Newton} observation will provide us with a better understanding of this system: with the iron K lines, the redshift of the components will be constrained, and cluster residual emission will be modelled with higher precision. This will allow us to understand if the different components are part of the same supercluster structure or not.

The global properties of the galaxy groups studied in this work are in good agreement with measurements for a sample of groups at lower redshift ranges and with scaling relations fitted to those samples. The main differences between our derived fluxes and luminosities, either from ROSAT or {\it XMM-Newton}, may be explained by the distinct techniques and data used.

We have found that $>1\%$ of $\sim 240$ eeHIFLUGCS clusters are the result of projection effects. This percentage represents a firm lower limit to the fraction of contaminated clusters. Further work is needed in order to assess the contamination fraction of full eeHIFLUGCS sample. This will help to determine the level of contamination of low-resolution catalogues, like those that will be available from eROSITA.

In summary, our results show that low-angular resolution X-ray surveys can produce biased cluster catalogues due to cluster projection effects. This bias is three-fold: fluxes are biased high, cluster numbers are biased low, redshifts are biased high or low. In this work, we show that with X-ray follow up using higher spatial resolution instruments we can identify multiple cluster components. We also demonstrate that given sufficient spectral resolution and collected X-ray photons, X-ray redshifts can robustly separate projected components. In general, X-ray spectral analysis represents an independent mean to determine the redshift also for less peculiar clusters in low-angular resolution X-ray cluster catalogues. This implies that with {\it XMM-Newton}, {\it Chandra}, and XRISM follow-up of subsamples of clusters to be discovered with eROSITA we will be able to quantify and correct for those biases.

\begin{acknowledgements}
MERC, FP and THR acknowledge support by the German Aerospace Agency (DLR) with funds from the Ministry of Economy and Technology (BMWi) through grant 50 OR 1514. KM is supported by the International Max Planck Research School (IMPRS) for Astronomy and Astrophysics at the Universities of Bonn and Cologne and the Bonn-Cologne Graduate School (BCGS) of Physics and Astronomy. L.L. acknowledges support from the {\it Chandra} X-ray Center through NASA contract NNX17AD83G. GS acknowledges support by the National Aeronautics and Space Administration (NASA) through \textit{Chandra} Award Number GO4-15129X issued by the \textit{Chandra} X-ray Observatory Center (CXC), which is operated by the Smithsonian Astrophysical Observatory (SAO) for and on behalf of NASA under contract NAS8-03060. Preliminary work on one of the targets was done by University of Bonn student Vyoma Muralidhara. This research has made use of: the NASA/IPAC Extragalactic Database (NED) which is operated by the Jet Propulsion Laboratory, California Institute of Technology, under contract with the National Aeronautics and Space Administration; and the VizieR catalogue access tool, CDS, Strasbourg, France.
\end{acknowledgements}

\bibliographystyle{aa}
\bibliography{references}

\begin{appendix}

\section{Detailed analysis of RXCJ0034.6-0208}
\label{app:numberA}

For the analysis of the RXCJ0034.6-0208 cluster system we use the extracted spectra in circles of $3$~arcmin radius centred on each cluster component. The quality of the data does not allow us to detect the iron K or L complex in either of the structures. We fitted the spectra for each detector separately as explained in Sec.~\ref{sub:spectralanalysis}. We report the results for the redshift analysis in Table~\ref{tab:apendixA}.

\begin{figure}[hb]
\centering
\includegraphics[width=\columnwidth]{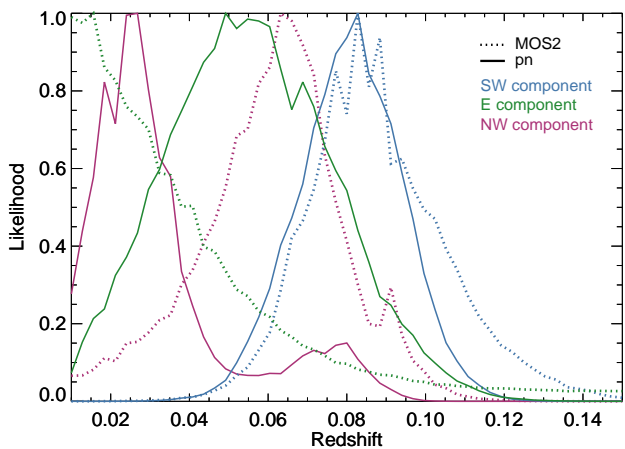}
\caption{\footnotesize{Variation of $\mathcal{L}$ as a function of the redshift for the different extended components in the RXCJ0034.6-0208 observation. Solid and dotted lines represent the MOS2 and pn detectors, respectively; the blue, green and purple lines show the results for the southwestern, eastern and northwestern components, respectively.}}
\label{fig:logLvariation}
\end{figure}

The redshift measurements for each detector in the southwestern component are consistent within $1\sigma$. A joint fitting of all available instruments gives a consistent redshift result ($z=0.082$, see Sec.~\ref{subsect:epzTZ}). For the northwestern structure, the result of redshift estimation with MOS2 is higher than the results with pn, although they are consistent at the $3\sigma$ level. In this case, a joint fit gives $z=0.028$, indicating that the pn spectrum determines the redshift estimate, which is not surprising given the much higher sensitivity of pn. In a similar way, the redshift results of pn of the eastern component settle the final value of the joint fitting ($z=0.045$) since the values of MOS2 cannot be constrained. Figure~\ref{fig:logLvariation} shows the variation of $\mathcal{L}$ for the individual fits of the three components.

The redshift discrepancies between the detectors suggest that there is a systematic problem in the observation. This is supported by the lack of MOS1 data in the observation.

\begin{table}[hb]
\centering
\caption{Redshift estimations from the X-ray spectral analysis for the components of RXCJ0034.6-0208.}
\label{tab:apendixA}
\begin{threeparttable}
\renewcommand{\arraystretch}{1.35} 
  \begin{tabular}{l c c}
   \hline
   \hline
    Components & MOS2 &pn \\
   \hline
     SW & $0.086^{+0.007}_{-0.015}$ & $0.081 ^{+0.012}_{-0.012}$ \\
     E & -- & $0.055^{+0.020}_{-0.020}$ \\
     NW & $0.064^{+0.005}_{-0.008} $ & $0.026^{+0.006}_{-0.010}$ \\
   \hline
   \hline
  \end{tabular}
\end{threeparttable}
\end{table}

\end{appendix}

\end{document}